\documentclass[graphicx,11pt]{aastex}
\ifx\pdfoutput\@undefined\usepackage[usenames,dvips]{color}
\else\usepackage[usenames,dvipsnames]{color}
\IfFileExists{pdfcolmk.sty}{\usepackage{pdfcolmk}}{} 
\fi

\lefthead{}
\righthead{}

\begin{document}

\title{Considerations on the magnitude distributions of the Kuiper
  belt and of the Jupiter Trojans}

\author{\textbf{Alessandro Morbidelli}}
\affil{\small\em Observatoire de la C\^ote d'Azur\\
  Boulevard de l'Observatoire\\
  B.P. 4229, 06304 Nice Cedex 4, France}

\author{\textbf{Harold F$.$ Levison}}
\affil{\small\em Southwest Research Institute\\ 
1050 Walnut St, Suite 300\\ 
Boulder, CO 80302~~~USA}
%\authoremail{hal@boulder.swri.edu}

\author{\textbf{William F$.$ Bottke}}
\affil{\small\em Southwest Research Institute\\ 
1050 Walnut St, Suite 300\\ 
Boulder, CO 80302~~~USA}

\author{\textbf{Luke Dones}}
\affil{\small\em Southwest Research Institute\\ 
1050 Walnut St, Suite 300\\ 
Boulder, CO 80302~~~USA}

\author{\textbf{David Nesvorn{\' y}}}
\affil{\small\em Southwest Research Institute\\ 
1050 Walnut St, Suite 300\\ 
Boulder, CO 80302~~~USA}

\newpage

\begin{abstract}

By examining the absolute magnitude ($H$) distributions (hereafter HD)
of the cold and hot populations in the Kuiper belt and of the Trojans
of Jupiter, we find evidence that the Trojans have been captured from
the outer part of the primordial trans-Neptunian planetesimal disk. We
develop a sketch model of the HDs in the inner and outer parts of the
disk that is consistent with the observed distributions and with the
dynamical evolution scenario known as the `Nice model'. This leads us
to predict that the HD of hot population should have the same slope of
the HD of the cold population for $6.5<H<9$, both as steep as the slope of
the Trojans' HD. Current data partially support this prediction, but
future observations are needed to clarify this issue.  Because the HD
of the Trojans rolls over at $H\sim 9$ to a collisional equilibrium
slope that should have been acquired when the Trojans were still
embedded in the primordial trans-Neptunian disk, our model implies
that the same roll-over should characterize the HDs of the Kuiper belt
populations, in agreement with the results of Bernstein et al. (2004)
and Fuentes and Holman (2008). Finally, we show that the constraint on
the total mass of the primordial trans-Neptunian disk imposed by the
Nice model implies that it is unlikely that
the cold population formed beyond 35 AU.

\end{abstract}

\section{Introduction}
\label{inro}

Models have recently proposed that the Jovian Trojan asteroids and the
Kuiper belt objects share a common origin --- they formed in a massive
primordial disk that stretched from roughly 15 to $\sim\!30\,$AU and
were transported to their current locations { during the
phase of orbital migration of the giant} planets (Morbidelli et
al$.$~2005; Levison et al$.$~2008).  This connection is a main result
of the so-called {\it Nice model} (Tsiganis et al., 2005; Gomes et
al., 2005).  In the Nice model, the giant planets are assumed to have
formed in a compact configuration (all were located between
5--$15\,$AU) and be surrounded by a $\sim\!35 M_\oplus$ planetesimal
disk that extended to about $30\,$AU.  It introduced the idea that
Jupiter and Saturn were so close in the past that they had to migrate
across their mutual 1:2 resonance. This led to a violent, but
temporary phase of instability in the dynamics of the four outer
planets.  The gravitational interaction between the ice giants and the
planetesimals damped the orbits of these planets - leading them to
evolve onto their current orbits.

As a result, however, $\sim\!35 M_\oplus$ of planetesimals were
scattered throughout the Solar System.  Some were then captured into
stable orbits by the migrating planets.  In particular, as Jupiter and
Saturn passed through various resonances with one another, a small
number of planetesimals would have been captured into the Trojans
regions of Jupiter.  Morbidelli et al$.$~(2005) showed that this
process quantitatively reproduces both the number and the orbital
element distribution of the observed Trojan swarms.  Another example
can be found in the trans-Neptunian region, which, according to
Levison et al.~(2008; L08 hereafter), was populated during the
high-eccentricity phase of Neptune.  The { simulations} in
L08 are the most successful to date at reproducing the observed
characteristics of the Kuiper belt.

If the above argument is correct, then there should be a genetic link
between the Trojan asteroids of Jupiter and the Kuiper belt objects
(hereafter KBOs), which should be detectable by studying their
physical characteristics.  Perhaps the most significant physical
property of a population is its size-distribution, or equivalently,
for a size-independent albedo, its absolute magnitude ($H$)
distribution (hereafter HD).  In particular, if the Trojan and Kuiper
belt populations are related they should have similar
HDs. This assumes that they have not undergone significant
collisional evolution since they were emplaced in their current
orbits.  Levison et al$.$~(2008b) has shown that this is a reasonable
assumption for the Trojans. The fact that, as we show below, the HDs
of the Trojans and the Kuiper belt are consistent with one another
argues that this is a reasonable assumption for the Kuiper belt as well
--- at least at the sizes we are concerned with here.  Thus, the
goal of this paper is to study the HDs of the Trojans and the Kuiper
belt to search for any genetic link.

Before we proceed, however, we need to discuss the structure of the
Kuiper belt in more detail.  The Kuiper belt has very intriguing
properties that show that the primordial disk of trans-Neptunian
planetesimals was drastically sculpted by a variety of dynamical
processes. A characteristic of particular relevance here is the
co-existence of {\it cold} and {\it hot} populations (defined by
having inclination respectively smaller and larger than 4.5 degrees;
Brown, 2001) with different physical properties (Levison and Stern,
2001; Tegler and Romanishin, 2000, 2003; Doressoundiram et al., 2001,
2005; Trujillo and Brown, 2002; Bernstein et al., 2004, B04 hereafter;
Elliot et al., 2005; see however Pexinho et al., 2008 for a
proposed alternative inclination divide between populations of
different color properties).  Levison and Stern (2001) showed that all of the
largest KBOs are found in the hot population.  This led them to
suggest that the hot population formed closer to the Sun (where large
objects would form more quickly) than the cold population, and were
transported outward as the orbits of the planets evolved (see
Gomes, 2003).  The difference in the size-distributions was confirmed
by B04, who showed that the HD for the hot population is shallower at
the bright end than that of the cold.

L08 explained the differences between the hot and cold populations in
the context of the Nice model by showing that the cold population is
derived almost exclusively from the outer part of the disk, while the
hot population samples the full disk more evenly. Thus they concluded
that the different HDs of the cold and the hot populations at the
bright end can be explained if the HD was not uniform throughout the original
planetesimal disk; instead, the outer part of the disk had a steep HD
and the inner part had a shallow HD and contained the largest
objects.

Studies of the collisional accretion and erosion of planetesimals
(e.g. Kenyon and Bromley, 2004) show that a small-body population
should have a cumulative {\it size} distribution that can be exemplified by
a broken power-law: the exponent $q_1$ of this power-law for large
sizes is a characteristic of the accretion process and thus it can
have different values for different populations; the exponent $q_2$
for small sizes is instead quite universal and characteristic of
collisional equilibrium (i.e. $q_2\sim -2$). The roll-over of the
power-law from exponent $q_1$ to exponent $q_2$ occurs around a size
$R_{\rm break}$ which can also differ from one population to
another. Remember now that, if the albedo is size-independent, a
cumulative power-law size distribution with exponent $q$ is equivalent
to an exponential HD of type $N(<H)\propto 10^{\alpha H}$, with
$\alpha=-q/5$. Thus, it is reasonable to expect that the inner and
outer parts of the original trans-Neptunian disk each followed a
broken HD, with their own values of $\alpha_1$ at the bright end
($\alpha_1^{(i)}$ for the inner disk, $\alpha_1^{(o)}$ for the outer disk,
with $\alpha_1^{(i)}<\alpha_1^{(o)}$) and rolling over to $\alpha_2\sim
0.4$ at some $H\sim H_{\rm break}$. {These broken
  HDs with different $\alpha$-slopes can be visualized in
  Fig.~\ref{model}.}

The Nice model implies that the HD of the Trojans, hot Kuiper belt,
and cold Kuiper belt populations should each be a combination of the
HDs found in the inner and outer parts of the original planetesimal
disk. In practice, in each magnitude range, these populations should
have inherited the HD of the part of the disk from which they captured
most of the objects. Thus we expect that the currently observed
Trojans and Kuiper belt populations have HDs with three exponents: the
brightest objects ($H<H_T$) should have $\alpha=\alpha_1^{(i)}$; the
intermediate objects ($H_T<H<H_{\rm break}$) should have
$\alpha=\alpha_1^{(o)}$; the faint end ($H>H_{\rm break}$) should have
$\alpha=\alpha_2$.  The absolute magnitude at the transition between the
first two exponents, $H_T$, is a function of the mixing ratio of the
inner and outer disk in each population and therefore could change
from one population to another.  In our analysis below, we assume this
simple HD as a template for analyzing the various populations of interest.

%we will assume that $H_T$ is the same for the Trojans
%and hot Kuiper belt because the simulations in Morbidelli et al.
%(2005) and L08 show that these populations result from a mixture of
%the disk {(This is not true BTW)}, while $H_T$ is smaller for the
%cold population because it comes primarily from the outer disk.

\section{Absolute magnitude distributions of multi-opposition objects}

We have considered the list of multi-opposition trans-Neptunian
objects given by the Minor Planet Center (see
http://www.cfa.harvard.edu/iau/lists/TNOs.html) as of April 14, 2008.
This list excludes the so-called Scattered disk objects (Duncan and
Levison, 1997; Luu et al., 1997). We have selected the bodies with
$40<a<47.4$~AU, in order to exclude those in the 2:3 and 1:2
resonances with Neptune, to simplify the discussion. Of the selected
objects, those with $i<4.5^\circ$ have been classified in the cold
population, and the remainder in the hot population, in agreement with
previous studies (Brown, 2001; Trujillo and Brown, 2002; Elliot et
al., 2005).

Fig.~2 shows the cumulative HDs in the cold (solid blue curve) and
hot (solid cyan curve) populations.  It is evident that the HD in the
hot population appears to be much shallower. This, however, could be
due to a bias, because part of the hot population has been discovered
in wide field surveys with shallower limiting magnitudes that strayed
farther from the ecliptic than the deep surveys that discovered the
smaller cold population objects.  Thus, in Fig.~2, we also plot the HD of the
hot population objects discovered within $4.5^\circ$ (solid purple
curve) from the ecliptic. In doing this, we believe that, although we
have not removed all observational selection effects, we have chosen
our objects so that the hot and cold population suffer from similar
biases.  Thus, we believe that any difference that we see in the
distributions are real.  Still, the hot HD appears shallower than
that of the cold population.  As we explained above, this result is
not new  (Levison and Stern 2001; B04).

In Fig.~2 the observed low-$H$ end (i.e. large sizes) of the HD of
the cold population is fit with a line of slope $\alpha=1.1$ (dashed
blue line). This slope is somewhat shallower than the preferred slope
($1.3$) of B04, but falls within its 1-$\sigma$ uncertainty (which
extends down to $\alpha \sim 1$). In addition, we find that a line
with the best fit slope of the debiased hot population in B04 (i.e.
$\alpha = 0.65$, dashed purple line) matches the observed HD of the
low-latitude hot population reasonably well. This general agreement
with B04 argues that the slopes of the observed low-latitude HDs that
we determined with our simple techniques { from the MPC
  catalogue} do not suffer from
significant observational biases.  Thus, we feel comfortable comparing
our results directly to the Trojans.

As a first step in our analysis, we compare the HDs of the cold
population with that of the Trojans of Jupiter.  In order to make this
comparison, however, we need to scale the observed distributions in a
physically meaningful way.  Using the information in B04, we estimate
that there should be between about 50 and 200 objects in the cold
population with $H<6$.  The dotted blue curves in Fig.~3 show the
observed cold population scaled to these two extremes.  To scale the
Trojans, we make use of the fact that the Nice model predicts that
both the cold Kuiper belt and the Trojans come from the primordial
trans-Neptunian disk and provides the corresponding capture
efficiencies.  In particular, L08 showed that between 0.04\% and
0.16\% of the full original disk (inner plus outer parts) was captured
in the cold Kuiper belt population during the migration of the
planets.  Morbidelli et al$.$~(2005) calculated that the fraction of
the full disk that was captured into the Trojans swarms with low
libration amplitude orbits was between $0.85\times 10^{-7}$ and
$0.65\times 10^{-6}$.  Thus, we expect that the ratio of cold KBOs to
Trojans should be between 600 and 19,000.  The dotted green curves in
the figure show the Trojan HD (which is complete to $H=11$; Szab\'o et
al., 2007) scaled by these factors.

The solid blue and green curves in Fig$.$~3 show the observed HDs of
the cold KB and the Trojans, respectively, scaled using factors that
are consistent with the constraints discussed in the last
paragraph. They illustrate that the two distributions have essentially
the same slope at the bright end (we attribute the roll-off in the
cold population for $H > H_{\rm bias} \sim 6.7$ to observational bias).

Of course, this close similarity between the slopes could be
coincidental. We think, however, that this is unlikely, because this
common slope is very steep and peculiar.  Indeed, in the asteroid
belt, no sub-population has a similar HD (with the possible exception
of some super-catastrophic families). The cold population/Trojan slope
is very far from a collisional equilibrium slope ($\alpha=$0.4--0.5; Dohnanyi,
1969). It is also steeper than what a prolonged phase of collisional
coagulation would give (S.~Kenyon, private communication). Thus, it
would be odd that the Trojans and the cold population show the same
slope if they were unrelated. Instead, we believe that their common HD
slope is real and reveals the genetic link predicted by the Nice model
(see sect.~1). 

The Trojans' HD shows a roll-over at $H_{\rm break}\sim 9$.  Beyond
this magnitude, the HD of the Trojans has a shallower slope with
exponent $0.4$ (Jewitt et al., 2000; Szab\'o et al., 2007), consistent
with the expected value of $\alpha_2$ (see sect.~1).  Notice that the
Trojan population is observationally complete down to absolute
magnitude $H\sim 11$ (Szab\'o et al., 2007), so the observed roll-over
of its HD is a real feature.  Unfortunately, we do not know with
certainty the Kuiper belt HDs at these magnitudes because of its
greater distance.  However, if the arguments in sect.~1 are correct,
both the hot and cold populations Kuiper belt should show the same
type of roll-over at these magnitudes. This is consistent with the
results of B04 and Fuentes and Holman (2008).

We interpret the common slope of Trojans and cold KBOs
as the slope of the HD of the outer part of the primordial
trans-Neptunian disk (i.e. $\alpha_1^{(o)}=1.1$).  The lack of a shallow
part in the Trojans' HD (i.e. with $\alpha\sim \alpha_1^{(i)}$)
indicates that the $H_T$ for this population is brighter than the
largest Trojan, i.e. $H_T\lesssim 7.5$.  Similarly, the lack of a shallow part
in the HD of the cold Kuiper belt suggests that this population must
have $H_T\lesssim 4.8$.  

We now turn to the hot population.  As we described in sect.~1, we
expect that there is a value of $H_T$, characteristic of this
population, such that $\alpha=\alpha_1^{(i)}$ for $H<H_T$ and
$\alpha=\alpha_1^{(o)}$ for $H_T<H<H_{\rm break}$. Because
$\alpha_1^{(o)}$ is also the slope of the cold population, we expect
that the HD of the hot population to steepen up so that its slope
matches that of the cold population for objects fainter than $H_T$.

In Fig.~4 we show the same green (scaled Trojan) and blue (cold KB)
curves as were in Fig$.$~3, but we added the observed HD of the hot
population discovered within $4.5^\circ$ of the ecliptic. We choose
this subset of the hot population so that the observational biases are
comparable for the two populations. To make the comparison more
straightforward, we multiplied the hot population curve by a factor of
2.5.  As one sees, the hot population HD is shallower for $H\lesssim
6.5$--7.0, but overlaps almost perfectly the HD of the cold population
beyond this magnitude threshold. So, our prediction seems to be
confirmed and $H_T$ for the hot population turns out to be $\sim 6.7$.
{This is not in conflict with Bernstein et
al. (2004), who claimed that the the luminosity functions of the hot and cold
populations have significantly different slopes for apparent magnitude
$R\lesssim 23$ because, for an average distance of 42~AU, $R=23$ is
equivalent to $H=6.7$.}

A closer inspection of Fig.~4, however, reveals an important and
troubling issue.  {Although the HDs of the hot and cold
populations appear indistinguisheable for $H\gtrsim 6.7$, the HD of the
hot population does not seem to become steeper, unlike what we would
expect. Instead, it remains constant from $H\sim 6.0$ to $H\sim
7.5$. It is the HD of the cold population that becomes shallower to
match that of the hot population! Above, we have argued that the
reason for this apparent deviation of the cold population's HD from the
$\alpha=1.1$ line at $H\sim 6.7$ is that $H_{\rm bias}\sim 6.7$.
If this is true, then Fig.~4}  implies that $H_T \sim H_{\rm
bias}$ and the bias conspires with the increase in the hot
population's $\alpha$ to produce a constant slope. This would be an
amazing coincidence. To avoid this coincidence, the alternative
interpretation is that $H_{\rm bias}\sim 7.5$ and that the HD of the hot
belt has really a single slope up to at least $H_{\rm bias}$.  This,
though, would imply that the apparent roll-over of the HD of the cold
population at $H\sim 6.7$ is also real (because biases change the
slope only for $H>H_{\rm bias}\sim 7.5$). We see two problems with
this alternative interpretation. First, it would seem another amazing
coincidence that, whatever physical process caused this roll-over
\footnote{presumably collisions, although it has never been shown that
collisions could flatten the HD of bodies down to $H=6.5$ which,  
depending on albedo, corresponds to diameter
$D=250$--300~km.}, it gave to the cold population exactly the same
slope that the hot population has at larger sizes. Second, a roll-over
at $H\sim 6.5$ would be in conflict with the results of all surveys
devoted to unveil the true slope of the Kuiper belt luminosity
function (Gladman et al., 2001; B04; Petit et al., 2006; Fraser et
al., 2008; Fuentes and Holman, 2008), which reported no roll-over up
to at least $H\sim 8$--9.  On the other hand, it is also true that no
team has ever reported a steepening of the HD of the hot population, but
there have never been dedicated pencil beam surveys off ecliptic, so
this feature, extended only over a couple of magnitudes, might have
passed un-noticed.

Thus, for the moment we prefer our original interpretation to its
alternative, although we feel quite uncomfortable about the
coincidence that it implies. {However, this is a testable hypothesis.
If the above argument is correct, we predict that the
hot population's HD becomes as steep as the common
cold-population/Trojan slope ($\alpha\sim 1.1$) for $H$ in the $\sim$6.5-9
interval}. These predictions can be checked in the future when more
data are collected and observational biases are removed.

\section{The mass of the trans-Neptunian disk}

The Nice model requires that the original mass of the trans-Neptunian
planetesimal disk was 35--50~$M_\oplus$ (Tsiganis et al., 2005; Gomes
et al., 2005).  It is not obvious, a priori, that this requirement is
consistent, even at the order of magnitude level, with the size
distributions that we can infer from the considerations reported in
the previous section for the inner and the
outer parts of the disk. Here we discuss this consistency check.

\begin{figure}[h!]
\centerline{\includegraphics[height=7.cm]{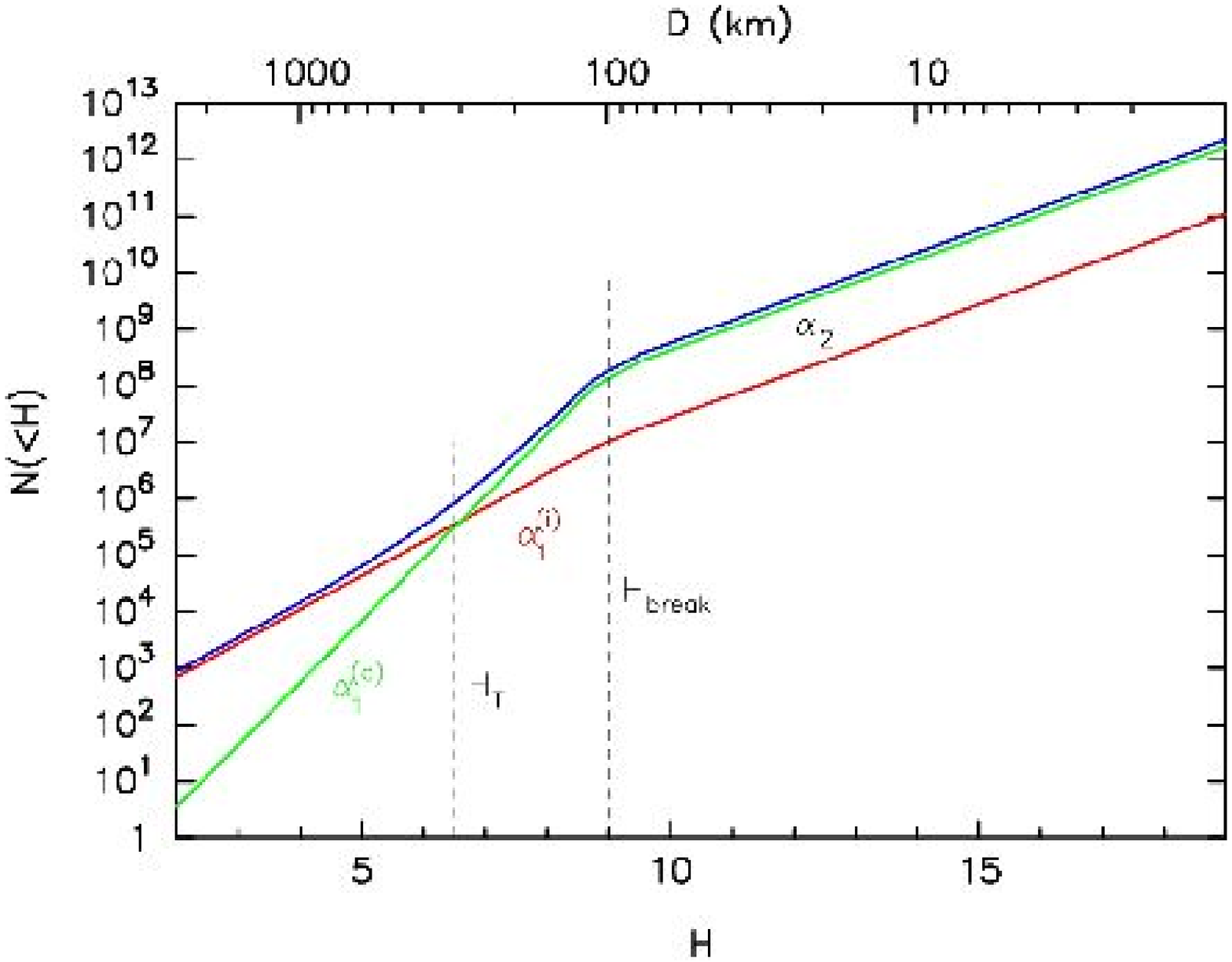}}
\vspace*{-.3cm}
\caption{ Sketch of the HDs in the inner (red) and outer
  (green) parts of the disk, showing the meaning of the quantities
  $\alpha_1^{(i)}$, $\alpha_1^{(o)}$, $\alpha_2$, $H_T$ and $H_{\rm
  break}$. The values of these quantities that we adopt to draw this
  figure are those that characterize the disk model defined in
  Sect.~3. The blue curve shows the HD obtained summing the HDs for
  the inner and the outer parts of the disk. We expect that the
  current HDs of the cold and of the hot populations of the Kuiper
  belt have the same shapes of the green and of the blue curves,
  respectively. The HD of the Trojans has the shape of the blue curve,
  but is defined only for $H>7.5$ (the absolute magnitude of the
  brightest Trojan). The correspondence between the size scale (top)
  and $H$ scale (bottom) has been obtained assuming a 4.5\%
  albedo. For simplicity, we assume that the albedo is
  size-independent.}
\label{model}
\end{figure}

In section~2 we concluded that $H_T$ for the hot population is $\sim
6.5$. The simulations in L08 show that { particles in the
  inner and outer parts of the disk have roughly the same probability 
 of being captured in the hot population. So, the value of $H_T$ 
characterizing the hot population should have been also the one 
at which the HDs of the
inner and the outer parts of the disk crossed-over (i.e. $N_{\rm
  inner}(<H_T)=N_{\rm outer}(<H_T)$; see Fig.~\ref{model}).}

We have also seen that the the cold population has 50-200 objects with
$H=6$, which is equivalent to 175 to 700 objects with $H<H_T\equiv
6.5$, for $\alpha=1.1$. The fraction of the full trans-Neptunian disk
captured in the cold population is 0.04\%--0.16\% (L08). Thus, there
were between $10^5$ and $1.7\times 10^6$ objects with $H<H_T$ in the
disk, half of which in the inner and in the outer parts. Assuming an
albedo of 4.5\%, $H=6.5$ corresponds to $D\sim 300$~km.

For the outer part of the disk, $\alpha_1^{(o)}=1.1$ implies that the exponent
of the cumulative size distribution is $q_1=-5.5$. The size of the
largest object in the outer disk is the value $D_{\rm max}$ such that
$N(>D_{\rm max})$=1. Given $q_1$ and the estimated number range of 300~km
objects obtained above, $D_{\rm max}$ is 2,100-3,500~km.

For the inner part of the disk, we assume that the size distribution
is truncated at $D_{\rm max}=$2,500~km (Pluto-size) and $q_1=-3$
(i.e. $\alpha_1^{(i)}=0.6$, which is within the slope uncertainties
for the hot population; see B04). Scaling again from the number 
of 300~km objects, this gives between $\sim$ 100 and 1,500 objects of
size $D_{\rm max}$. This is consistent with the estimate (L08) that
roughly 1,000 Pluto-size objects had to exist in the disk (Charnoz and
Morbidelli, 2007) estimated this number to be $\sim 300$).

For both the inner and the outer disks we assume $H_{\rm break}=9$. 
Assuming again that the albedo is 4.5\%, this corresponds to $D_{\rm
  break}=100$~km. For smaller objects, we assume $\alpha_2=0.4$
(i.e. $q_2=-2$) as for the Trojans. 

Now we have all the ingredients to compute the total mass (we assume a
bulk density of 1g/cm$^3$). We find that the mass of the inner disk is
between 1.5 and 23 Earth masses; the mass of the outer disk is between
9 and 130 Earth masses. The total mass of the full disk is between 10
and 150 Earth masses. The order of magnitude uncertainty on the total
mass comes from the comparable uncertainty on the total number of
objects, illustrated by the green curves in Fig.~3. Nevertheless, the
fact that our estimate brackets the disk mass required in the
dynamical simulations of Tsiganis et al. (2005) and Gomes et
al. (2005) gives, once again, confidence on the gross consistency of
that model.

{ The extremes of the mass estimate correspond to $N(D>300{\rm
km})=10^5$ and $1.7\times 10^6$, respectively. The total mass scales
almost linearly with $N(D>300{\rm km})$, so that a total mass of 35
Earth masses would imply $N(D>300{\rm km})\sim 4\times 10^5$. The HDs
shown in Fig.~\ref{model} have been normalized to this value.}

\section{Cold population: local or implanted?}

In the original version of the Nice model, the planetesimal disk is
assumed to be truncated at $\sim 34$~AU (Tsiganis et al., 2005; Gomes
et al., 2005). The Kuiper belt is therefore empty. The cold population
is captured into the Kuiper belt from within 34~AU (L08). 

However, Morbidelli et al. (2008) showed that, from a purely dynamical
point of view, the Nice model is also consistent with the existence
of a local, low mass Kuiper belt population, extended up to 44~AU. In
fact, the resulting $(a, e)$ distribution of cold population objects
would be indistinguishable from that in L08.  In the case where the
outer edge was initially at 44~AU, about 7\% of the particles initially
in the Kuiper belt ($a > 40$ AU, $q > 38$ AU) remained there, although
on modified orbits. The others escaped to planet-crossing orbits
during the large eccentricity phase of Neptune. Thus, a
local population could be consistent with the Nice model provided that
its total number of objects in the 40--44~AU interval was about 15
times the one currently present in the cold population.

Here we examine this possibility at the light of the considerations of
the previous sections. In particular, we re-work our estimates
assuming that the outer disk contained 15 times the current number of
bodies with $D>300$~km in the cold population in the 40--44~AU
interval (i.e. from 2,500 to 10,000 bodies) and show that this would
imply a radial distribution of the disk's surface density that is
unlikely. 

Our argument is the following.
Given that the median initial semi major axis of the bodies trapped in
the Trojan region is $\sim 26$~AU, we assume that this was the
boundary between the inner and the outer parts of the disk. If the
surface density of the disk had a radial profile as $1/r$, there would
have been an equal number of objects per linear AU. In this case, as
2,500--10,000 bodies with $D>300$~km were spread over 4~AU (40--44~AU
interval), the full outer disk (26--44~AU) would have contained
11,000--45,000 bodies. Assuming a $1/r^2$ surface density profile, the
total number would increase to 14,000--55,000.  Remember now that the
total number of bodies larger than 300~km (or $H<6.5\equiv H_T$) should have
been the same in the inner and in the outer parts of the disk (because
of the very definition of $H_T$).  As we
have seen in the previous section, if this number had really been
$\lesssim$ 50,000, the total mass of the full disk would have been
$\lesssim 10 M_\oplus$, too low for the dynamics of the Nice
model. Moreover, the disk would have contained less than $5\times
10^5$ bodies larger than 200~km (assuming $q_1^{(o)}=-5.5$), too low
to explain the capture of a few Trojans of this size given a capture
probability lower than $6.5\times 10^{-7}$ (Morbidelli et al., 2005).

To have a total mass in the disk of $\sim 35$ Earth masses, 
the number of $D>300$~km bodies in each part
of the disk should have been $\sim$~200,000, as computed in
the previous section. With 2,500--10,000 bodies in the 40--44~AU region, this
could have been achieved only if the disk had a surface density
profile of $r^{-\beta}$ with $\beta\sim 7$--11.  This is much steeper
than any radial profile ever proposed or inferred from observations of
extra-solar disks.

Therefore we conclude that, in the framework of the Nice model, it is
unlikely that the planetesimal disk extended into the Kuiper belt
region. Thus, the cold population should have been implanted into the
Kuiper belt from a smaller initial heliocentric distance.

\section{Conclusions}

In this paper, we have given a fresh look at the HDs of the Kuiper
belt objects and of the Jovian Trojans.  We have partitioned the Kuiper belt
objects between 40 and 47 AU into a cold and a hot population,
according to their orbital inclinations (Brown, 2001).  We have
confirmed that the HD of the hot population is shallower than that of
the cold population (Fig.~2) for $H<6.5$ (B04), which supports models
in which these two populations are derived from different regions of
the primordial trans-Neptunian planetesimal disk. 

We have found that the slope of the bright end of the Trojan
population is very similar to that of the cold Kuiper belt population
(Fig.~3).  This, if not just a coincidence, suggests a genetic link
between the two populations. Of all the models proposed so far on the
history of the Kuiper belt, only the Nice model predicts such a link.

Thus, we have developed a sketch model of the HDs in the inner and the
outer parts of the trans-Neptunian disk that is consistent with the
aspects of the Nice model and explains the similarity of the HDs of
the Trojans and of the cold population. This HD model has led us to
predict that the HD of the hot population should steepen up, so that
its slope matches that of the cold population for $H\gtrsim 6.5$. The
current data seem to support this prediction because they show that
the HDs of the hot and of the cold populations are identical beyond
this magnitude threshold (Fig.~4). However, we pointed out that it is
disturbing that the {\it observed} HD of the hot population has a straight
slope up to $H\sim 7.5$, because this would imply that the
steepening up of the real HD is perfectly counterbalanced by the
observational biases. Thus, this issue remains to be settled with
future observations.

Our HD model of the disk also implies that it is unlikely that the
cold population formed in situ, but suggests that this population was
implanted into the Kuiper belt from a smaller heliocentric distance.

\vskip 20pt
{\centerline {\bf Acknowledgments}}

This work was done while the first author was on sabbatical at
SWRI. A.M. is therefore grateful to SWRI and CNRS for providing the 
opportunity of this long term visit and for their financial support.

\clearpage

\centerline{\bf References}

\begin{itemize}

%\item[$\bullet$] Allen, R.~L., Bernstein, 
%G.~M., Malhotra, R.\ 2001.\ The Edge of the Solar System.\ Astrophysical 
%Journal 549, L241-L244. 
%
%\item[$\bullet$] Allen, R.~L., Bernstein, 
%G.~M., Malhotra, R.\ 2002.\ Observational Limits on a Distant Cold Kuiper 
%Belt.\ Astronomical Journal 124, 2949-2954. 

\item[$\bullet$] Bernstein, G.~M., 
Trilling, D.~E., Allen, R.~L., Brown, M.~E., Holman, M., Malhotra, R.\ 
2004.\ The Size Distribution of Trans-Neptunian Bodies.\ Astronomical 
Journal 128, 1364-1390. 
%
%\item[$\bullet$] Bottke, W.~F., Levison, 
%H.~F., Morbidelli, A., Tsiganis, K.\ 2008.\ The Collisional Evolution of 
%Objects Captured in the Outer Asteroid Belt During the Late Heavy 
%Bombardment.\ Lunar and Planetary Institute Conference Abstracts 39,
%1447. 

\item[$\bullet$] Brown, M.~E.\ 2001.\ The 
Inclination Distribution of the Kuiper Belt.\ Astronomical Journal 121, 
2804-2814. 

\item[$\bullet$] Charnoz, S., 
Morbidelli, A.\ 2007.\ Coupling dynamical and collisional evolution of 
small bodies. II. Forming the Kuiper belt, the Scattered Disk and the Oort 
Cloud.\ Icarus 188, 468-480. 

\item[$\bullet$]Dohnanyi, J.~W.\ 1969.\ 
Collisional models of asteroids and their debris.\ Journal of Geophysical 
Research 74, 2531-2554. 

\item[$\bullet$] Doressoundiram, 
A., Barucci, M.~A., Romon, J., Veillet, C.\ 2001.\ Multicolor Photometry of 
Trans-neptunian Objects.\ Icarus 154, 277-286. 

\item[$\bullet$] Doressoundiram, 
A., Peixinho, N., Doucet, C., Mousis, O., Barucci, M.~A., Petit, J.~M., 
Veillet, C.\ 2005.\ The Meudon Multicolor Survey (2MS) of Centaurs and 
trans-neptunian objects: extended dataset and status on the correlations 
reported.\ Icarus 174, 90-104. 

\item[$\bullet$] Duncan, M.~J., 
Levison, H.~F.\ 1997.\ A scattered comet disk and the origin of Jupiter 
family comets.\ Science 276, 1670-1672. 

\item[$\bullet$] Elliot, J.~L., and 10 
colleagues 2005.\ The Deep Ecliptic Survey: A Search for Kuiper Belt 
Objects and Centaurs. II. Dynamical Classification, the Kuiper Belt Plane, 
and the Core Population.\ Astronomical Journal 129, 1117-1162. 
%
%\item[$\bullet$] Fern{\'a}ndez, 
%Y.~R., Sheppard, S.~S., Jewitt, D.~C.\ 2003.\ The Albedo Distribution of 
%Jovian Trojan Asteroids.\ Astronomical Journal 126, 1563-1574. 

\item[$\bullet$] Fraser, W.~C., 
Kavelaars, J.~J., Holman, M.~J., Pritchet, C.~J., Gladman, B.~J., Grav, T., 
Jones, R.~L., Macwilliams, J., Petit, J.-M.\ 2008.\ The Kuiper belt 
luminosity function from $m_{R}=21$ to 26.\ Icarus 195, 827-843. 

\item[$\bullet$] Fuentes, C.~I., 
Holman, M.~J.\ 2008.\ a SUBARU Archival Search for Faint Trans-Neptunian 
Objects.\ Astronomical Journal 136, 83-97. 

\item[$\bullet$] Gladman, B., Kavelaars, 
J.~J., Petit, J.-M., Morbidelli, A., Holman, M.~J., Loredo, T.\ 2001.\ The 
Structure of the Kuiper Belt: Size Distribution and Radial Extent.\ 
Astronomical Journal 122, 1051-1066. 

\item[$\bullet$] Gomes, R.~S.\ 2003.\ The origin 
of the Kuiper Belt high-inclination population.\ Icarus 161, 404-418. 

\item[$\bullet$] Gomes, R., Levison, 
H.~F., Tsiganis, K., Morbidelli, A.\ 2005.\ Origin of the cataclysmic Late 
Heavy Bombardment period of the terrestrial planets.\ Nature 435, 466-469. 
%
%\item[$\bullet$] Grundy, W.~M., Noll, 
%K.~S., Stephens, D.~C.\ 2005.\ Diverse albedos of small trans-neptunian 
%objects.\ Icarus 176, 184-191. 
%
%\item[$\bullet$] Hahn, J.~M., 
%Malhotra, R.\ 2005.\ Neptune's Migration into a Stirred-Up Kuiper Belt: A 
%Detailed Comparison of Simulations to Observations.\ Astronomical Journal 
%130, 2392-2414. 

\item[$\bullet$] Jewitt, D.~C., Trujillo, 
C.~A., Luu, J.~X.\ 2000.\ Population and Size Distribution of Small Jovian 
Trojan Asteroids.\ Astronomical Journal 120, 1140-1147. 
%
%\item[$\bullet$] Kobayashi, H., Ida, 
%S., Tanaka, H.\ 2005.\ The evidence of an early stellar encounter in 
%Edgeworth Kuiper belt.\ Icarus 177, 246-255. 

\item[$\bullet$] Kenyon, S.~J., 
Bromley, B.~C.\ 2004.\ The Size Distribution of Kuiper Belt Objects.\ 
Astronomical Journal 128, 1916-1926. 

\item[$\bullet$] Levison, H.~F., 
Stern, S.~A.\ 2001.\ On the Size Dependence of the Inclination Distribution 
of the Main Kuiper Belt.\ Astronomical Journal 121, 1730-1735. 
%
%\item[$\bullet$] Levison, H.~F., 
%Morbidelli, A.\ 2003.\ The formation of the Kuiper belt by the outward 
%transport of bodies during Neptune's migration.\ Nature 426,  

\item[$\bullet$] Levison, H.~F., 
Morbidelli, A., Vanlaerhoven, C., Gomes, R., Tsiganis, K.\ 2008.\ Origin of 
the structure of the Kuiper belt during a dynamical instability in the 
orbits of Uranus and Neptune.\ Icarus 196, 258-273. 

\item[$\bullet$] Levison, H.~F., Bottke, 
W., Gounelle, M., Morbidelli, A., Nesvorny, D., Tsiganis, K.\ 2008b.\ 
Chaotic Capture of Planetesimals into Regular Regions of the Solar System. 
II: Embedding Comets in the Asteroid Belt.\ AAS/Division of Dynamical 
Astronomy Meeting 39, \#12.05. 
%
%\item[$\bullet$] Lykawka, P.~S., 
%Mukai, T.\ 2007.\ Origin of scattered disk resonant TNOs: Evidence for an 
%ancient excited Kuiper belt of 50 AU radius.\ Icarus 186, 331-341. 
%
%\item[$\bullet$] Lykawka, P.~S., 
%Mukai, T.\ 2008.\ An Outer Planet Beyond Pluto and the Origin of the 
%Trans-Neptunian Belt Architecture.\ Astronomical Journal 135, 1161-1200. 

\item[$\bullet$] Luu, J., Marsden, B.~G., 
Jewitt, D., Trujillo, C.~A., Hergenrother, C.~W., Chen, J., Offutt, W.~B.\ 
1997.\ A New Dynamical Class in the Trans-Neptunian Solar System..\ Nature 
387, 573-575. 

\item[$\bullet$] Morbidelli, A., 
Levison, H.~F., Tsiganis, K., Gomes, R.\ 2005.\ Chaotic capture of 
Jupiter's Trojan asteroids in the early Solar System.\ Nature 435, 462-465. 

\item[$\bullet$] Morbidelli, A., 
Levison, H.~F., Gomes, R.\ 2008.\ The Dynamical Structure of the Kuiper 
Belt and Its Primordial Origin.\ The Solar System Beyond Neptune 275-292. 

\item[$\bullet$] Peixinho, N., Lacerda, 
P., Jewitt, D.\ 2008.\ Color-Inclination Relation of the Classical Kuiper 
Belt Objects.\ Astronomical Journal 136, 1837-1845. 

\item[$\bullet$] Petit, J.-M., Holman, 
M.~J., Gladman, B.~J., Kavelaars, J.~J., Scholl, H., Loredo, T.~J.\ 2006.\ 
The Kuiper Belt luminosity function from $m_{R}= 22$ to 25.\ Monthly Notices 
of the Royal Astronomical Society 365, 429-438. 

\item[$\bullet$] Szab{\'o}, G.~M., 
Ivezi{\'c}, {\v Z}., Juri{\'c}, M., Lupton, R.\ 2007.\ The properties of 
Jovian Trojan asteroids listed in SDSS Moving Object Catalogue 3.\ Monthly 
Notices of the Royal Astronomical Society 377, 1393-1406. 

\item[$\bullet$] Tegler, S.~C., 
Romanishin, W.\ 2000.\ Extremely red Kuiper-belt objects in near-circular 
orbits beyond 40 AU.\ Nature 407, 979-981. 

\item[$\bullet$] Tegler, S.~C., 
Romanishin, W.\ 2003.\ Resolution of the Kuiper belt object color 
controversy: two distinct color populations.\ Icarus 161, 181-191. 
%
%\item[$\bullet$] Trujillo, C.~A., 
%Brown, M.~E.\ 2001.\ The Radial Distribution of the Kuiper Belt.\ 
%Astrophysical Journal 554, L95-L98. 

\item[$\bullet$] Trujillo, C.~A., 
Brown, M.~E.\ 2002.\ A Correlation between Inclination and Color in the 
Classical Kuiper Belt.\ Astrophysical Journal 566, L125-L128. 

\item[$\bullet$] Tsiganis, K., Gomes, 
R., Morbidelli, A., Levison, H.~F.\ 2005.\ Origin of the orbital 
architecture of the giant planets of the Solar System.\ Nature 435, 
459-461. 

\end{itemize}

\clearpage

\begin{figure}[h!]
\centerline{\includegraphics[height=7.cm]{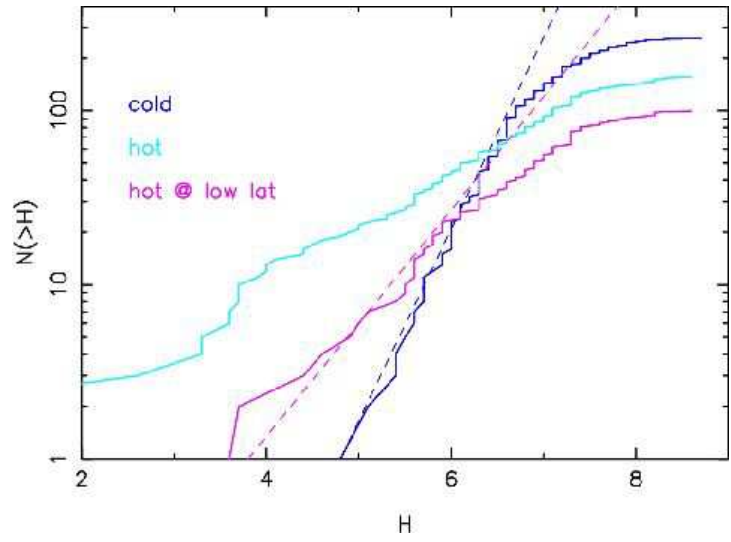}}
\vspace*{-.3cm}
\caption{{The absolute magnitude ($H$) cumulative distributions of the
  cold population (blue), of the hot population (cyan) and of the
  objects of the hot population discovered within $4.5^\circ$ of the
  ecliptic (magenta). For reference, the thin dashed blue line has a
  slope of $1.1$; the thin dashed magenta line has a slope of
  $0.65$.}}
\end{figure}

\begin{figure}[h!]
\centerline{\includegraphics[height=7.cm]{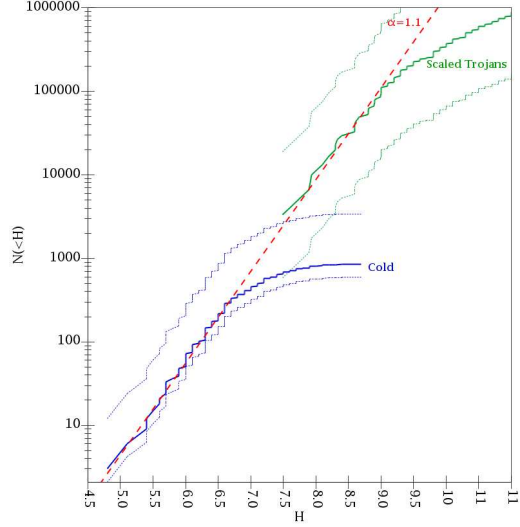}}
\vspace*{-.3cm}
\caption{The cumulative $H$-distributions of the cold population (blue
  lines) and of the Trojan population (green lines). The blue dotted
  lines bracket the current number of cold KBOs according to B04. The
  dotted green curves show the current HD of the Trojans, scaled by
  factors of 600 (lower curve) and 19,000 (upper curve), which
  correspond to the minimal and maximal ratios between the populations
  of cold KBOs and Trojans, according to the Nice model. The solid
  blue and green curves show the observed HDs of the cold KBOs and
  Trojans, scaled by appropriate factors so that they fall within the
  boundaries provided by the dotted curves. The red dashed line
  corresponds to an HD with $\alpha=1.1$. This line highlights that
  the HD of the Trojan population with $H<9$ has the same slope of the
  HD of the cold population with $H<6.5$. We interpret the apparent turn-over
  of the HD of the cold population at $H=6.5$ as due to observational
  biases. The common slope of the HDs of cold KBOs and Trojans
  supports the genetic link between these two populations predicted by
  the Nice model.}
\end{figure}

\begin{figure}[h!]
\centerline{\includegraphics[height=7.cm]{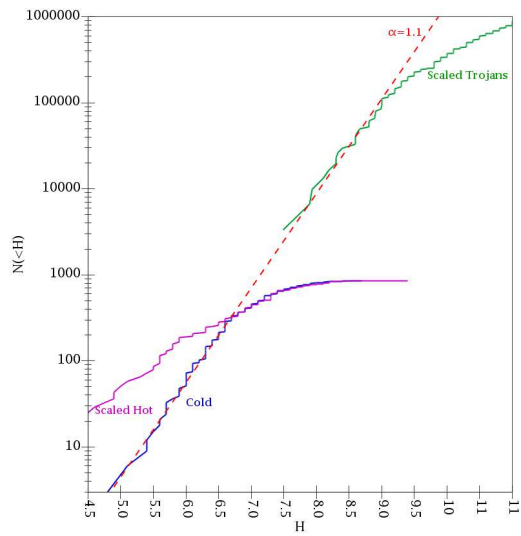}}
\vspace*{-.3cm}
\caption{The same as Fig. 2, but only the solid curves for the cold
  KBOs and Trojans are plotted, while the HD of the hot population has
  been added, scaled up by a factor of 2.5. Notice that the observed
  HDs of the cold and of the hot populations are identical for
  $H>6.5$, in agreement with  our predictions. See text for a broader
  discussion on this point.}
\end{figure}

\end{document}